\documentclass[article]{elsarticle}
\usepackage{lineno,hyperref}
\usepackage{gensymb}
\modulolinenumbers[5]
\usepackage{tabularx} 
\usepackage{graphicx}
\usepackage{setspace}

\journal{Journal of \LaTeX\ Templates}

\begin{document}

\begin{frontmatter}

\title{An improved approach to manufacture CNT reinforced magnesium AZ91 composites with increased strength and ductility}

\author{Samaneh Nasiri$^1$, Feng Shi$^2$, Guang Yang $^3$, \\ Erdmann Spiecker$^3$, Qianqian Li$^{*,4}$}

\address{$^1$Chair of Materials Simulation, Department of Materials Science, Friedrich-Alexander Universit\"at Erlangen-N\"urnberg, 90762 F\"urth, Germany}	
\address{$^2$Institute of Advanced Materials and Processes (ZMP), Friedrich-Alexander Universit\"at Erlangen-N\"urnberg, 90762 F\"urth, Germany}
\address{$^3$ Chair of Micro- and Nanostructure Research, Friedrich-Alexander Universit\"at Erlangen-N\"urnberg, 91058 Erlangen, Germany}
\address{$^4$Department of Aeronautics, Imperial College London, UK} 
\cortext[mycorrespondingauthor]{Corresponding author: Qianqian Li, Tel. +44 2075945109; Fax. +44 2075945056; Email: Qianqian.Li@imperial.ac.uk }

\begin{abstract}

Multiwalled carbon nanotubes (MWCNTs) are decorated with Pt nanoparticles by a "layer-by-layer" approach using poly (sodium 4-styrene sulfonate) (PSS) and poly (diallyl dimethylammonium chloride) (PDDA). Transmission electron microscopy (TEM) images and Energy Dispersive X-Ray (EDX) analysis of the samples confirm Pt deposition on surfaces of CNTs. Dispersibility and dispersion stability of MWCNTs in the solvents are enhanced when MWCNTs are coated with Pt nanoparticles. Mg AZ91 composites reinforced with MWCNTs are then produced by a melt stirring process. Compression tests of the composites show that adding 0.05\% wt Pt-coated MWCNTs in AZ91 improves the composite's mechanical properties compared to the pure AZ91 and pristine MWCNT/AZ91. Fracture surface analysis of the composite using a scanning electron microscope (SEM) shows individuals pulled out MWCNTs in the case of the Pt-coated MWCNT/AZ91 composites. We attribute this finding to the uniform dispersion of Pt-coated MWCNTs in Mg due to the improved wettability of Pt-coated MWCNTs in Mg melts. Molecular dynamics (MD) simulations of the interaction between Pt-coated MWCNTs and Mg support this interpretation.

\end{abstract}


\end{frontmatter}

\onehalfspacing
\section{Introduction}
\label{intro}

Carbon nanoparticle (CNPs) in particular carbon nanotubes (CNTs) and graphene flakes (GPs) with excellent mechanical properties \cite{Sammalkorpi2004-PRB, Khare2007-PRB, Lee2008-science, nasiri2016-AIMSmaterialsscience}, high specific surface area \cite{zhang2013-Sci.Rep, stoller2008-Nanoletters, Zhang2020-mat.lett}, and enhanced thermal properties \cite{Deng2014-App.Phys.lett,pop2012-MRSbulletin,balandin2011-Nat.Mater} are promising candidates as reinforcement agents in lightweight metals such as Al and Mg \cite{Esawi2008-Comp.sci.technol,Bakshi2011-carbon}.

The benefits of embedding well-dispersed carbon nanoparticles into metals are potentially huge, as the embedded CNTs can carry part of the load and increase effective elastic moduli of the composite\cite{Nam2012-carbon,Chen2017-Acta,Chen2015-CST,Li2009-cops.sci.technol}. It is also shown that CNTs can bridge incipient cracks \cite{Chen2015-CST, Chen2019-Jalcom} and impose a strong constraint on dislocation motion \cite{kim2013-Nature, Nasiri2022-materialia, Chang2013-Philos.Mag.Lett}, potentially increasing yield strength, fracture strength, and composite toughness \cite{choi2012-NanoEnergy, George2005-scripta, Li2009-PRL}. Molecular dynamics simulation (MD) of tension and compression of metal matrices reinforced with CNTs and GPs confirmed the experimental observations and report an increase in Young's modulus, fracture stress and plasticity of the metal matrices due to the presence of CNPs \cite{Choi2016-JCOMB, Park2018-JMST, Xiang2017-commatsci,Nasiri2022-materialia-crack}. The importance of interfacial properties between CNPs and metal matrix, arrangement of CNPs inside the matrix and CNPs length  on the tensile strength and deformation behaviour of the composite and crack propagation patterns inside the matrix are discussed in details in \cite{Nasiri2022-materialia-crack}.  

To which extent the potential benefits of embedding CNP into metals can be harnessed in practice is a mostly open question. Experimental studies indicate that low-melting lightweight metals such as Al or Mg do not wet $sp^2$ bonded carbon, as Zhang et al. \cite{Zhang2011-Mat.Chem.Phys} showed that the wetting angle between the basal plane of graphite and molten Mg is about 120\degree. Consequently, immersed CNPs into the molten metals tend to agglomerate while the interfacial bonding between the nanoparticles and the matrix remains poor after solidification. Goh et al. \cite{Goh2005-nanotechnology,Goh2006-Mat.sci.eng.A} reported a decrease in the ductility of the Mg/CNT composite when a higher percentage of CNTs added in the pure Mg matrix, due to the fact that  agglomerated CNP with weak bonding into the surrounding metal matrix might act as flaws that deteriorate the mechanical properties rather than improve \cite{Zhou2007-composite.A,Deng2007-Mat.Sci.Eng.A,Li2010-compos.sci.technol}. 


One approach to overcome the CNP agglomerations in molten metals is inspired by earlier research on improving the wettability of graphite in Al melts and matrices by Ni coating of GPs \cite{Ip1998-Mater.Sci.Eng.A, Chu2000-wear, Guo2002-Mater.Sci.Eng.A}. Guo and Tsao \cite{Guo2002-Mater.Sci.Eng.A} reported that Ni coating of graphite particles improved graphite dispersion in the matrix. Chu and Lin \cite{Chu2000-wear} found superior wear properties for composites with Ni-coated graphite particles over composites with uncoated graphite fillers. Accordingly, the same strategy, i.e., coating the CNTs with an appropriate metal, looks promising in the context of metal-CNTs composites. The possibility of improving interfacial properties in CNT-metal systems by metal coating of CNTs has also been investigated by Song et. al. \cite{Song2010-comput-Mat-Sci,Duan2017-PhysicaE} using atomistic simulations to compare the pull-out behaviour of coated and uncoated CNTs from an Al matrix. These investigations show that Ni coating may significantly enhance the CNT pull-out force. Nasiri et al. \cite{Nasiri2019-EPJ} found that depending on the representation of interface between Ni and CNT, a huge pullout force up to 30 eV/$\mathrm{\AA}$ can be achieved in case of Ni-coated CNT, which exceeds the pull-out force of the uncoated CNT by almost two orders of magnitude and comes very close to the CNT fracture force. Ni-coated CNTs are also shown to hinder dislocation mobility in Al matrices stronger than un-coated CNTs, as the large elastic mismatch between Al and Ni increases the energy associated with a slip step in the surface of the cylinder surrounding the CNT interface\cite{Nasiri2022-materialia}. Coating CNTs with metal nanoparticles also has potentials applications in nanoelectronics, biosensors, and catalysis \cite{Kong2002-sct,jafri2009mno2,yang2006platinum}. Methods such as electrochemical deposition \cite{day2005electrochemical}, electroless deposition \cite{qu2005substrate}, and layer-by-layer coating \cite{du2009general} have been developed to this end.

Attaching metal nanoclusters rather than continuous metal coatings to CNT may also benefit composite properties. If a metal such as Pt is used that bonds well to $sp^2$ bonded carbon, such that the interfaces between the nanoclusters and the CNT are strong, then such nanoclusters may act as a ’nano-rivets’ enhancing interfacial shear stress transfer if the decorated CNT is embedded in a metal matrix. To understand this point it is important to note that small nanoparticles in themselves exhibit tremendous mechanical strength, as they are too small to contain dislocations. This is confirmed by experimental studies of Chen et al. \cite{Chen2019-Jalcom}, where presence of $\rm Al_2O_3$ nanoparticles at the Al/CNT interface showed an improvement in the interface shear stress (IFSS) between Al and CNT, which changes the failure behavior of CNTs in Al matrices from the pure CNT pullout (weak interface strength) to the CNT failure (strong interface shear strength). Nasiri et al. \cite{Nasiri2019-EPJ} demonstrate that metal nanoclusters deposited on the surface of CNT may act as efficient nano-crystallization agents and thus provide a novel strengthening mechanism. Besides, discrete metal nanoclusters on the CNP surface can prevent the agglomeration of CNPs by acting as geometrical spacers. Si et al. \cite{si2008-Chem.ofMat} showed that the decoration of exfoliated graphene sheets with Pt nanoclusters prevents face-to-face graphene aggregation. Atomistic simulation study of multilayer structure of graphene and Pt nanoparticle found an optimum nanoparticle density where the adhesive energy of the structure is lowest, and thus the separation of the graphene sheets most easy \cite{Nasiri2020-AEM}. 

In present work, we produce a Mg AZ91 composites reinforced with Pt-coated MWCNTs by a melt stirring process. A noble metal such as Pt is our choice as coating element because of its high chemical and thermomechanical stability with melting point of more than 2000K. Regarding the adhesion energy between Pt and $sp^2$ carbon bonds, density functional theory (DFT) calculations indicate larger adhesion energy in Pt/graphene compared to Al/graphene and Mg/graphene interface \cite{khomyakov2009-PRB, schneider2013-ChemPhysChem, ramos2013-Phys.Chem.Chem.Phys, gan2008-Small}. 

The coating approach through polyelectrolyte was chosen to coat MWCNTs as it was a fairly straightforward process and can be easily scaled-up. The Pt deposition on CNTs was studied by Transmission electron microscopy (TEM) images and Energy Dispersive X-Ray (EDX) analysis. We also investigated the dispersibility and dispersion stability of the coated CNTs in ethanol. We then produced metal-coated CNTs reinforced Mg AZ91 composites by a melt stirring process. The mechanical properties of Pt-coated CNT/AZ91 composite were then measured by compression testing. A scanning electron microscope (SEM) was used to analyze the fracture surface of the composites. To further understand the effect of Pt-coating on the mechanical properties of the composite, molecular dynamics (MD) simulations were performed to gain a fundamental understanding of the CNTs pull-out behavior during deformation.    

\section{Experimental procedure and Results}
\subsection{Synthesis of Mg AZ91 composite reinforced with Pt-coated MWCNTs} Pt coated MWCNTs were prepared using a modified layer-by-layer coating process\cite{du2009general}, in which 30 mg of MWCNTs (diameter of 5-20 nm, Baytubes$^{\textregistered}$ C 150P) were sonicated in 50 ml of 1 M NaCl solution (58.44 g of NaCl dissolved in 1 liter of distilled water) for 1 hour. 400 mg Poly(diallyldimethylammonium chloride) (PDDA) solution (Mw 200 000 - 350 000 Da, 20wt\% solution in water, SIGMA-ALDRICH) was added. The solution was stirred for 30 minutes. Then the excess PDDA was removed by six repeated centrifugation/wash cycles. Subsequently, 80 mg Poly(sodium 4-styrenesulfonate) (PSS, Mw 70 000 Da, Alfa Aesar) and 400 mg PDDA solution were added and removed similarly. The 30 mg polyelectrolyte-modified MWCNTs were put into 120 ml water with 36.8 mg of H2PtCl6 (Pt:MWCNTs = 3:7) and 240 mg of trisodium citrate dehydrate. The solution was sonicated for 30 minutes. Then 80 ml Na2SO3 (0.05 mol/l) was slowly dripped in under mild sonication. After 30 minutes, the resulting black solid product was centrifuged and washed with distilled water and ethanol to remove ions possibly remaining in the final product. The final product was dried at 80 \degree C in air.  

\begin{figure}
	\centering
	\resizebox{0.5\textwidth}{!}{
	\includegraphics{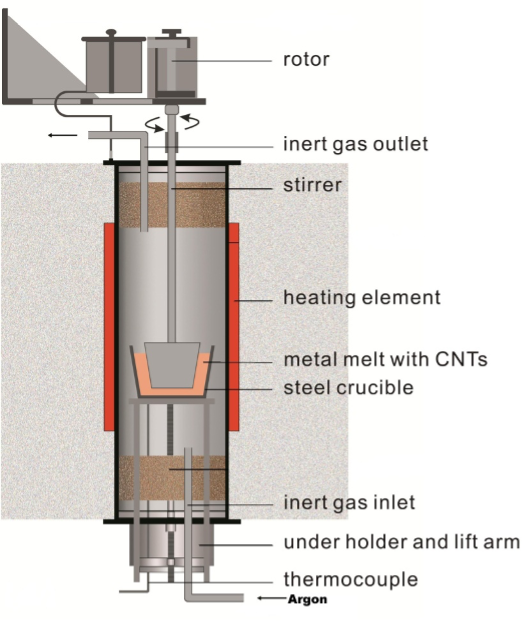}}
	\caption{Schematic drawing of the melt stirring furnace used in this study.}
\label{fig0}       
\end{figure}

The Pt-coated MWCNT/AZ91 composites were produced by a two-step process, including pre-dispersion and a melt stirring step as described in our previous work \cite{Li2009-cops.sci.technol, Li2010-compos.sci.technol, Hippmann2013-proc}. The block copolymer Disperbyk-2150 (BYK Chemie GmbH) was first dissolved in ethanol in a small beaker. Then Pt-coated MWCNTs (0.1 wt\% of the metal matrix, mass ratio to the block copolymer 1:1, diameter of 5-20 nm, Baytubes$^{\textregistered}$ C 150P) were added to the as-prepared solution. This mixture was put at room-temperature into an ultrasonic bath for 15 minutes. Then it was stirred for 30 minutes at 250 rpm. After adding Mg alloy chips (AZ91 D, ECKA), the suspension was further stirred at 250 rpm inside a fume cupboard to evaporate ethanol and homogenise the mixture. After the mixture was dried, the mixed chips were placed in a cylindrical sample crucible. This crucible was placed into an oven and heated up to 700 \degree C under an inert Argon gas atmosphere to avoid oxidation. When the Mg alloy chips were molten, the liquid was mechanically stirred at 350 rpm for 30 minutes to further disperse MWCNTs. After stirring, the molten MWCNT/Mg composite was poured into a mould. The cooled sample was machined to cylindrical shaped specimens (diameter 5 mm  height 7 mm) for subsequent compression tests. Non-coated MWCNT/AZ91 composite and pure AZ91 alloy were produced by using the same process for comparison. The schematic drawing of the melt stirring furnace used in this study is shown in Figure \ref{fig0}.

\subsection{TEM imaging and EDX analysis of Pt-coated MWCNTs} 

\begin{figure}
	\centering
	\resizebox{0.8\textwidth}{!}{
	\includegraphics{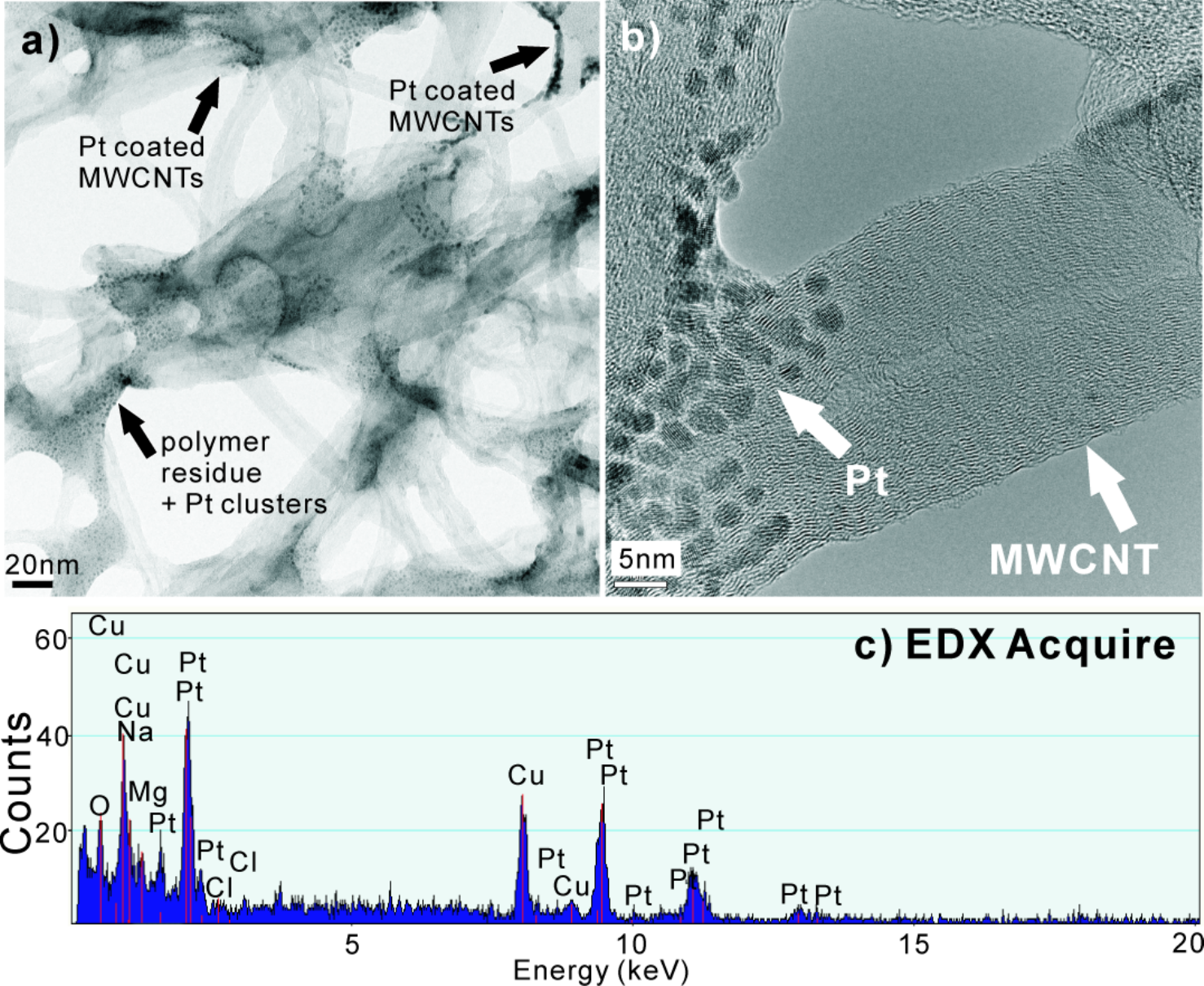}}
	\caption{TEM image of overview of Pt coated MWCNTs (containing 30 wt\% of Pt), containing Pt-coated MWCNTs and polymer residue with Pt clusters; (b) TEM image of Pt coating on the surface of individual MWCNTs; (c) EDX analysis of selected area on surface on Pt-coated MWCNTs.}
\label{fig1}       
\end{figure}

Transmission Electron Microscopy (TEM) images and EDX analysis of Pt-coated MWCNT are shown in Figure \ref{fig1}.  Figure  \ref{fig1} (a) confirms a uniform Pt nanoparticle distribution (black dots) on the surface of MWCNTs using our modified "layer-by-layer" coating approach. Polymer residue and Pt agglomerated clusters was still found on the surface of MWCNTs as pointed out by arrows. The higher magnification of TEM image (Figure \ref{fig1} (b)) shows that Pt have been coated as small clusters with diameters in range of 2.0 to 5.0 nm on the surface of the MWCNTs. EDX analysis confirmed Pt deposition on the surface of MWCNTs (Figure \ref{fig1} (c)).

\subsection{Dispersion of Pt coated MWCNTs} 
\begin{figure}
	\centering
	\resizebox{0.8\textwidth}{!}{
	\includegraphics{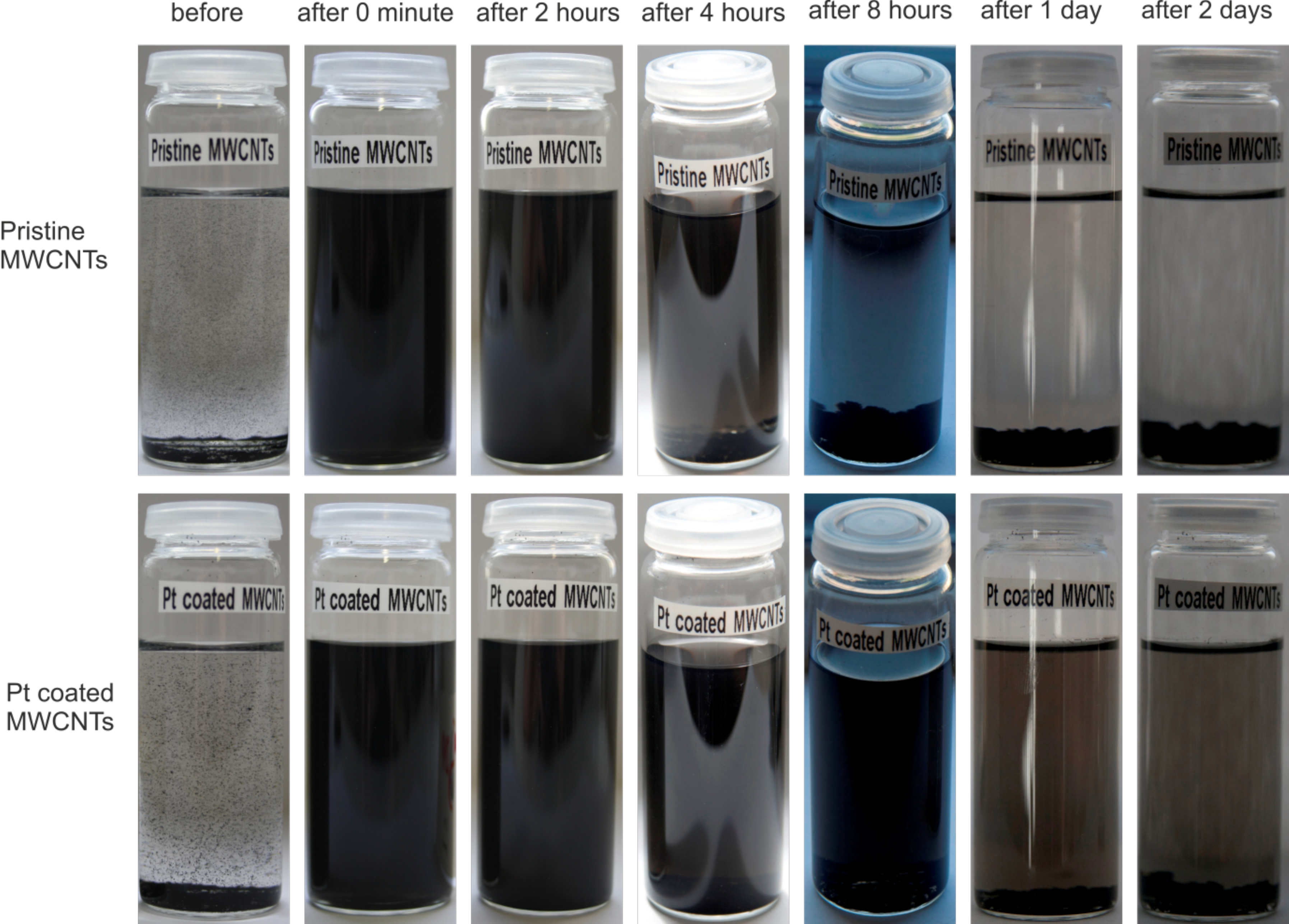}}
	\caption{Dispersion ability and stability of Pt coated MWCNTs and pristine MWCNTs.  Same weight (1.14 mg) of Pt coated MWCNT (coating ratio has been considered) and pristine CNTs in same amount of ethanol (20 ml). Sonicated for 15 minutes}
\label{fig2}       
\end{figure}

The dispersability and dispersion stability of Pt-coated MWCNTs were compared to pristine MWCNTs. By using the same amount of MWCNTs in the same amount of ethanol after the same treatment (ultrasonification for 15 minutes), the difference between Pt coated MWCNTs and pristine MWCNTs can be observed in Figure \ref{fig2}. Straight after ultrasonification, both solutions showed an evenly black colour (after 0 minute) and after two hours, the difference was still not too obvious. But after four hours, pristine MWCNTs started re-agglomerating and black MWCNT agglomerations can be observed on the bottom of the bottles (after 4 hours and after 8 hours); while for Pt-coated MWCNTs, the solutions were still black and no obvious agglomerations can be seen. After 1 day and 2 days, the solution of pristine MWCNTs showed a much lighter colour compared to Pt coated MWCNTs and big agglomerations can be found on the bottom; while the solution of Pt coated MWCNTs still presented a rather even black colour with fewer agglomerations on the bottom. The behaviour could be explained by that the Pt coating may act as spacers to break van der Waals force among the MWCNTs and to prevent them from re-agglomerating, therefore, the Pt coated MWCNTs have an improved dispersion ability and stability. 

\subsection{Compression tests of composite} 
\begin{figure}
	\centering
	\resizebox{0.6\textwidth}{!}{
	\includegraphics{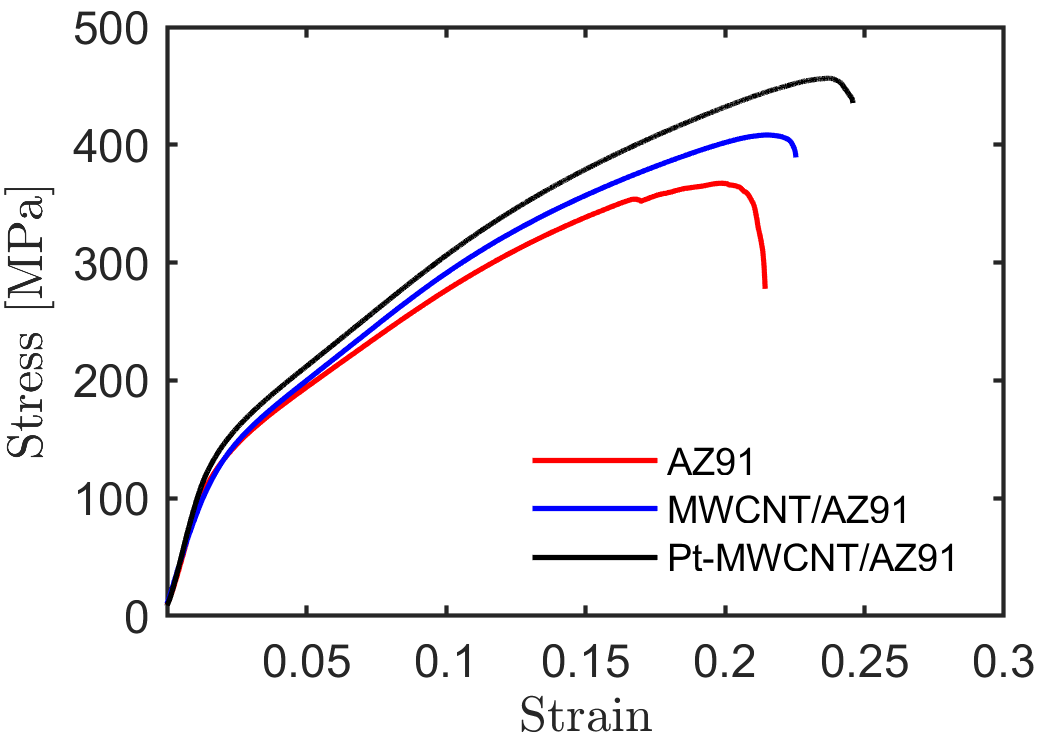}}
	\caption{Typical stress-strain curves of pure AZ91, 0.05 wt\% uncoated MWCNT/AZ91 composite and 0.05 wt\% Pt-coated MWCNT/AZ91 composite.}
\label{fig3}       
\end{figure}

We performed compression test to evaluated the effect of Pt coating on mechanical properties of the composite. To improve the statistical significance of the results, 36 specimens were tested following the same procedure. We compare the ultimate compressive strength (UCS), yield strength (YS), and compression strain at failure (CSF) for pure AZ91, MWCNT/AZ91 and Pt-coated MWCNT/AZ91 composites systems. The typical compression stress–strain curves in Figure \ref{fig3} illustrates that embedding MWCNTs and Pt-coated MWCNT can improve the mechanical properties of AZ91 composites.  

\begin{figure}
	\centering
	\resizebox{0.9\textwidth}{!}{
	\includegraphics{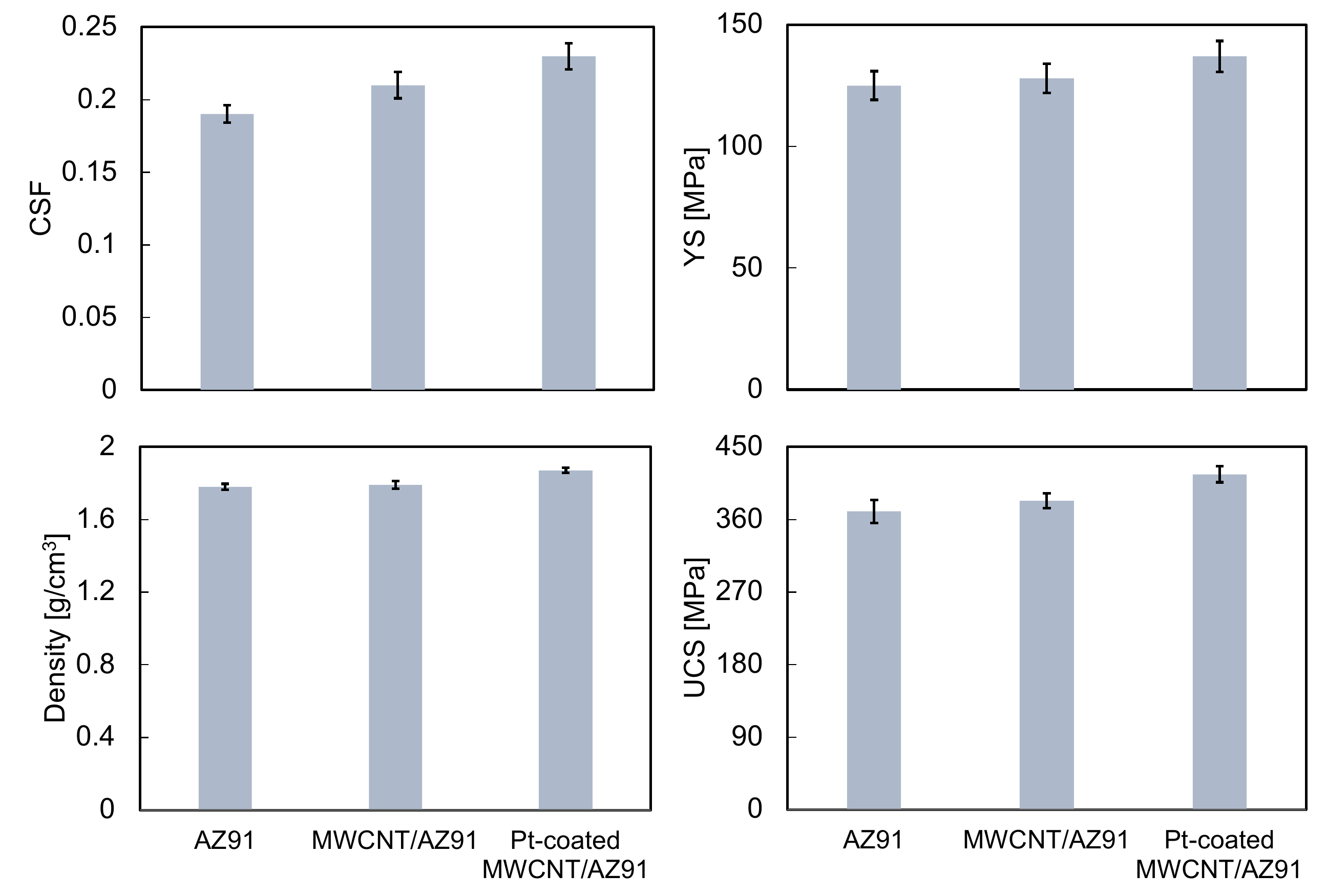}}
	\caption{ Comparison of (a) compression strain at failure (CSF), (b) yield strength (YS), (d) density, and (d) ultimate compressive strength (UCS) between pure AZ91, 0.05 wt\% MWCNT/AZ91 composite and 0.05 wt\% Pt-coated MWCNT/AZ91 composite.}
\label{fig4}       
\end{figure}

The yield strength of Pt-coated MWCNT/AZ91 has increased by about 10\% compared to pure AZ91 and 7\% compared to raw MWCNT/AZ91; Ultimate compressive strength of Pt-coated MWCNT/AZ91 has increased by 12\% compared to pure AZ91 and by 9\% compared to MWCNT/AZ91; Compression strain at failure of Pt-coated MWCNT/AZ91 has the most improvement as it increased by 21\% compared to pure AZ91 and by 10\% compared to raw MWCNT/AZ91. Density of the samples was also measured. By adding raw MWCNTs and Pt-coated MWCNTs, the densities of both composites have not been changed much compared to pure AZ91 alloy. A comparison between these values are depicted in Figure \ref{fig4} and detailed in Table \ref{tabl:LE}.

\begin{table}[]
	\centering
	\caption{Experimental values of yield strength (YS), ultimate compressive strength (UCS), compression strain at failure (CSF) and density of pure AZ91, 0.05 wt\% MWCNT/AZ91 and 0.05 wt\% Pt-coated MWCNT/AZ91.}
	\resizebox{\textwidth}{!}{
    \begin{tabular}{lllll}
	\hline\noalign{\smallskip}
	Samples& 2\% YS [MPa] & UCS [MPa]& CSF [\%] & Density [g/cm${^3}$]\\
	\noalign{\smallskip}\hline\noalign{\smallskip}
	AZ91&125 $\pm$ 5.6 & 370 $\pm$ 14.4  & 19 $\pm$ 0.6 & 1.78 $\pm$ 0.01\\
	MWCNT/AZ91 &128 $\pm$ 5.8 & 383 $\pm$ 9.3  & 21 $\pm$ 0.9 & 1.79 $\pm$ 0.02\\
	Pt-coated MWCNT/AZ91 &137 $\pm$ 7.0 & 416 $\pm$ 10.0  & 23 $\pm$ 0.9 & 1.87 $\pm$ 0.01\\
	\end{tabular}
	}	
	\label{tabl:LE}
\end{table} 

\begin{figure}
	\centering
	\resizebox{0.8\textwidth}{!}{
	\includegraphics{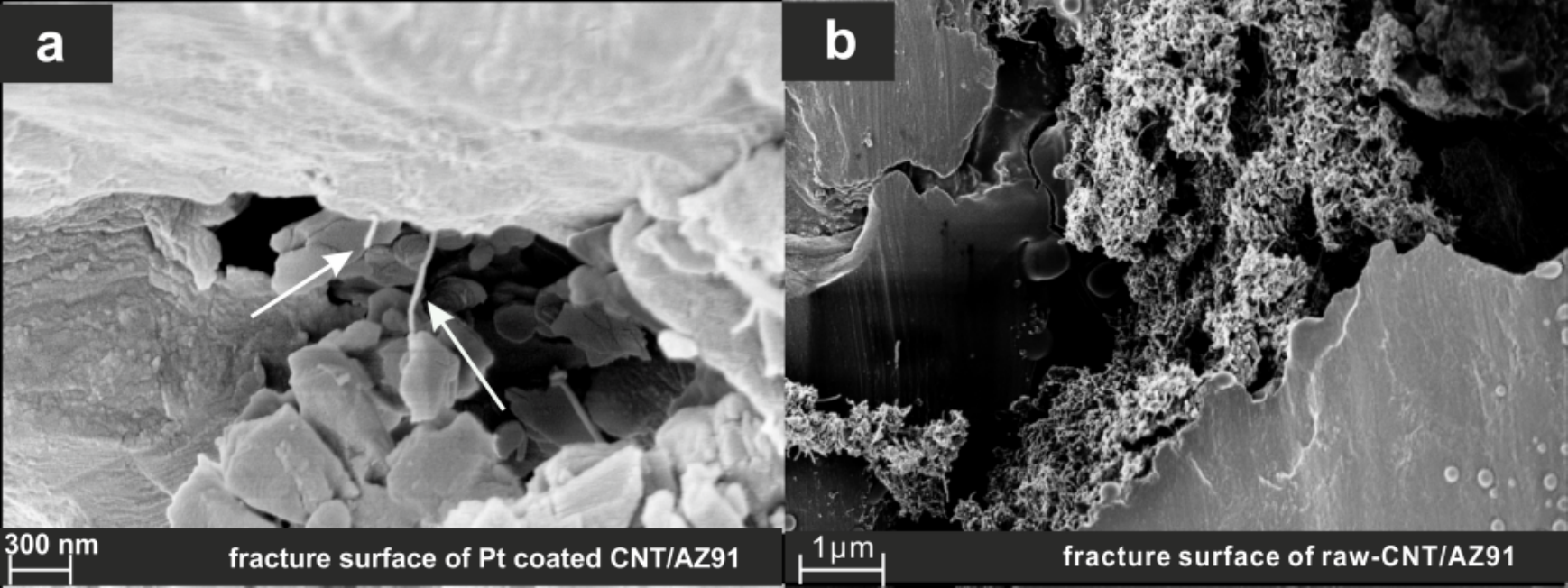}}
	\caption{Fracture surface of (a) Pt-coated MWCNT/AZ91 composite and (b) MWCNT/AZ91 composite after compression test. Arrow points at the individual MWCNT pulled out of the fracture surface after failure.}
\label{fig5}       
\end{figure}

Figure \ref{fig5} depicts the fracture morphology of Pt-coated and uncoated MWCNTs in Mg composites after compression testing characterized by scanning electron microscope (SEM).  In the case of uncoated MWCNT/AZ91 composite, we often observe micro-sized clusters of agglomerated MWCNT at the fracture surface as shown in Figure \ref{fig5}(b). The organic dispersion agent in uncoated MWCNT/Mg decomposes in the high temperature melting process. Eventually, in the absence of the organic dispersion agent and due to the MWCNT's high surface energy, large surface area, and poor wettability with Mg, MWCNTs tend to adhere to each other and agglomerate. On the other hand, in Pt-coated MWCNT/Mg composite, MWCNTs are observed to get pulled out individually at the fracture surface as shown in Figure \ref{fig5} (a). We attribute this observation to the fact that Pt nanoparticles deposited at the surface of MWCNT (are stable and do not melt away during the high temperature melting process) and act as spacer between the MWCNTs which inhibits the agglomeration of MWCNTs in Mg melt and matrices. We observe no fractured MWCNTs at the fracture surface of both Pt-coated and uncoated MWCNT/Mg composite. It means that the failure mechanism of MWCNT/Mg in both case of Pt-coated and uncoated MWCNTs are the pullout process. Attaching Pt nanoparticles on the surface of MWCNTs delays the pullout process of MWCNTs from the fracture surface as the interface energy and interface shear stress between Pt and CNTs are higher than Mg and CNT \cite{fampiou2012-J.Phys.Chem, schneider2013-ChemPhysChem, cheng2014-Acta}, which results in a higher ultimate compression strength and compression strain at failure as shown in Figure \ref{fig3} and Figure \ref{fig4}. To shed light on the pull-out behaviour of CNTs in Mg matrix, we perform pull-out simulations of Pt-coated and uncoated CNTs from Mg matrices using molecular dynamics simulation. 

\section{Atomistic simulation procedure and Results}

We consider a single crystal Mg block with faces aligned with the [1-210], [-1010], and [0001] directions. Mg atoms are initially put on a perfect hcp crystal with lattice constant of 3.19 $\mathrm{\AA}$. The simulation box of the Mg lattice is periodic in all directions. The energy of the system was minimized through structural relaxation using a CG algorithm. A single-walled CNT(5,5) is then embedded into the center of the Mg block, as  in our previous study, we demonstrated that for a constant diameter of CNT, increasing the number of CNT walls does not change the pull-out behaviour of MWCNTs \cite{Nasiri2019-EPJ}. To create the requisite space, we remove the Mg atoms that were located inside the CNT or within a distance of less than 3.5 $\mathrm{\AA}$ from carbon atoms. The longitudinal axis of the CNT is parallel to the [1-210], which aligns with $x$ direction of the Cartesian coordination's system. We carry out structural minimization using a CG algorithm and a thermal equilibration procedure, where Mg atoms were annealed at 1200K for 50 ps in $NpT$ ensemble at zero pressure, then quenched to 0.1K at a rate of 1 K/ps. A final structural minimization is then performed to remove any excess energy in the system.  In case of Pt-coated CNT/Mg composite, before embedding the Pt-coated CNT in Mg matrix the Pt coating layers around the CNT was annealed at 2300K for 20 ps in the $NpT$ ensemble at zero pressure, and then quenched to 0.1K at a rate of 10 K/ps. During this anneal-quench cycle, the CNT atoms are fixed. 

The standard AIREBO potential \cite{stuart2000-J.Chem.Phys.} was chosen for C-C interactions and the EAM/alloy potentials of Zhou et al. \cite{zhou2004misfit} for metal-metal interactions. The Mg-C interface was described using a standard 12-6 LJ potential ($\epsilon$ = 0.0027 eV and $\sigma$ = 3.75 $\mathrm{\AA}$) as in \cite{Zhou2016-J.Com.Mat}. A Morse potential the form $E_{\mathrm{M}} = D_{\mathrm{M}} (\exp[-2\alpha (r - r_0)] - 2 \exp[-\alpha (r - r_0)])$ was used for Pt-C interactions, where $r$ is the bond distance, $r_0$ the equilibrium bond distance, $D_\mathrm{M}$ is the well depth, and $\alpha$ controls the stiffness of the potential. The parameters for Pt-C morse potentials are $D_M$ = 0.0071 eV, $r_0$ = 4.18 $\mathrm{\AA}$ and $\alpha$ = 1.05 $\mathrm{\AA}^{-1}$ \cite{Nasiri2020-AEM}.

\begin{figure}
	\centering
	\resizebox{0.6\textwidth}{!}{
	\includegraphics{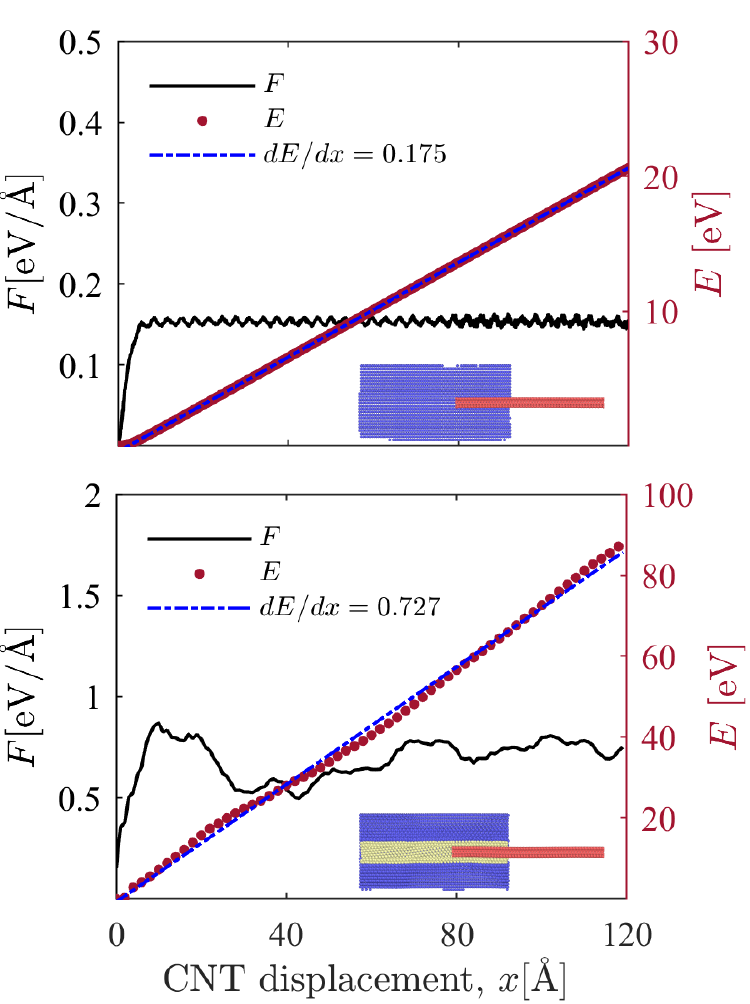}}
	\caption{MD simulation of CNT pull-out form Mg matrix; top: uncoated CNT in Mg matrix, bottom: Pt-coated in Mg matrix; inset pictures show the cross section of configuration during the pull-out simulation. Blue, red and yellow atoms at the insets represent Mg, C and Pt atoms, respectively.}
\label{fig6}       
\end{figure}

To set up a CNT pull-out test, the velocities of Mg atoms at the boundary of the simulation box were fixed to zero, and a constant velocity of 0.1 $\mathrm{\AA}$/ps was imposed for 1000 ps on the CNT front layer (30 Carbon atoms). The remaining atoms were thermostatted to a temperature of 1K using a Nos\'e -Hoover thermostat with a characteristic relaxation time of 0.1 ps. To have sufficient space in the pull-out direction, the simulation box was extended to more than twice the CNT length. During pull-out, the potential energy, as well as the total reaction force acting on the CNT front layer, were recorded.

Figure \ref{fig6} depicts the differences in pull-out force and energy between uncoated and Pt-coated (5,5) SWCNT. In both cases the pull-out behaviour are of pure sliding out without energy dissipation at the interface. This is consistent with our experimental observations in Figure \ref{fig5} when Pt-coated MWCNTs found pulled-out of the fracture surface. The pull-out force oscillates around a value of 0.175 eV/$\mathrm{\AA}$ for uncoated CNT and of 0.727 eV/$\mathrm{\AA}$ for Pt-coated CNT. These values match the average slope of the potentials energy increase which correspond to the product of CNT circumference and interface energy. The adhesion energy between Pt-C is higher than Mg-C, therefore, the pull-out force increases by a factor of five when CNT is coated by platinum. The corresponding interface shear stress (IFSS) between Mg and CNT is 10.9 MPa, and in the case of the Pt and CNT is 41.5 MPa. We should note that although the pull-out force and pull-out energy increases by Pt-coating, the interface characteristics between Pt and CNT is of physisorbed types in which no chemical bondings between metal and CNTs are formed. (\cite{Vanin2010-PRB, gong2010-J.Appl.Phys., fampiou2012-J.Phys.Chem}). The only commitment between metal and CNT in this model is the weak physical interactions, which will disappear under the shear stress of less than 50 MPa (\cite{Nasiri2020-AEM}). Therefore, CNTs, independent on their length, start to slide inside the matrix and get pulled out as soon as the matrix ruptures, and have little significant load-transfer strengthening effect on the Mg matrix.
  
\section{Conclusion}
Multiwall carbon nanotubes have been coated by island-like clusters of Pt atoms via polyelectrolyte PDDA and PSS. By adding small amounts of Pt coated MWCNTs, the mechanical properties of Mg AZ91 composites have been improved over pure AZ91, and also compared to composites produced using un-coated MWCNTs. We attributed this enhancement to a better dispersion of Pt-coated MWCNTs in AZ91 Mg melt and concluded that Pt-decoration may inhibit MWCNTs re-agglomeration in metal melt. In composites with Pt-coated MWCNTs, individual MWCNTs can be observed on the fracture surface indicating that MWCNTs were dispersed individually. Simulations demonstrated that Pt coating led to enhanced interfacial bonding between MWCNTs and Mg metal matrix. Future work will be needed to investigate the microstructure of the composites by TEM, to optimise the metal coating process, and to investigate whether the process can be transferred to different metal matrix and coating materials.  

\section{Acknowledgements}
Financial support by the German Research Foundation (DFG) under grant LI 1847/2-1 is gratefully acknowledged. The authors also gratefully acknowledge the support of the Cluster of Excellence 'Engineering of Advanced Materials' at the University of Erlangen-Nuremberg, which is funded by the German Research Foundation (DFG) within the framework of its 'Excellence Initiative’ under grant EXC 315-1. We thank Dr. Jan Schwerdtfeger, Claudia Backes and Christoph Dotzer for fruitful discussions. SN acknowledges DFG funding under Za171/11-1.

\bibliographystyle{unsrt}
\biboptions{sort&compress}
\bibliography{bibifile_MgPtC}

\end{document}